\documentclass[%
 reprint,
 amsmath,amssymb,
 aps,
]{revtex4-2}

\usepackage{graphicx}
\usepackage{dcolumn}
\usepackage{bm}
\usepackage{hyperref}
\usepackage{xurl}
\hypersetup{breaklinks=true}

\begin{document}

\preprint{APS/123-QED}

\title{Vorticity wave interaction and exceptional points in shear flow instabilities}

\author{Cong Meng$^{1,2}$ and Zhibin Guo$^{2}$}

 \email{zbguo@pku.edu.cn}

 \affiliation{$^{1}$
 Southwestern Institute of Physics, PO Box 432, Chengdu 610041, China}
 
\affiliation{$^{2}$
 State Key Laboratory of Nuclear Physics and Technology, Fusion Simulation Center, School of Physics, Peking University, Beijing 100871, China
}

\date{\today}

\begin{abstract}
We establish a link between vorticity wave interaction and $\mathcal{PT}$-symmetry breaking in shear flow instabilities. The minimal dynamical system for two coupled counter-propagating vorticity waves is shown to be a non-Hermitian system that exhibits a saddle-node exceptional point. The mechanism of phase-locking and mutual growth of vorticity waves is then related to the Krein collision and the breaking of $\mathcal{PT}$-symmetry through the exceptional point. The key parameter that leads the system to spontaneous $\mathcal{PT}$-symmetry breaking is the ratio between frequency detuning and coupling strength of the vorticity waves. The critical behavior near the exceptional point is described as a transition between phase-locking and phase-slip dynamics of the vorticity waves. 
The phase-slip dynamics lead to non-modal, transient growth of perturbations in the regime of unbroken $\mathcal{PT}$-symmetry, and the phase-slip frequency $\Omega \propto |k-k_c|^{1/2}$ shares the same critical exponent with the phase rigidity of system eigenvectors. The results can be readily extended to the interaction of multiple vorticity waves with multiple exceptional points and rich transient dynamics.

\end{abstract}

\maketitle


\section{\label{sec:intro}Introduction}

Interaction of vorticity waves has been widely demonstrated to be a physical interpretation of shear flow instabilities \cite{baines_1994,Heifetz_1999,heifetz_2004}, and naturally provides a non-modal approach to study optimal growth and transient dynamics in such instabilities \cite{Heifetz_2005,guha_2014,guha_2017}, as thoroughly reviewed in \cite{Carpenter_2011}. The model of coupled vorticity waves can be reframed into a minimal nonlinear dynamical system \cite{Heifetz_2019_PRE,Heifetz_PRE_2022}, where the onset of shear instabilities corresponds to a bifurcation of fixed points from two neutral centers to a pair of stable and unstable nodes \cite{Heifetz_2019_PRE}. The mechanism of instability onset is illustrated as phase-locking and mutual growth of the vorticity waves \cite{Carpenter_2011}.

On the other hand, the role of symmetries and symmetry-breaking bifurcations in fluid dynamics has long been recognized \cite{ARFM_1991,ARFM_2023}, and it has recently come to light that shear flow instabilities are the result of spontaneous $\mathcal{PT}$-symmetry breaking \cite{Hong_Qin_2019_PoP,Fu_2020_NJP,Tomos_David_2023}. In the framework of spectral analysis of pseudo-Hermitian, or equivalently, G-Hamiltonian systems \cite{Yakubovich1975,Kirillov2013,Kirillov2014}, the mechanism of instability onset is illustrated as the spontaneous breaking of $\mathcal{PT}$-symmetry and the Krein collision between eigenmodes with opposite Krein signatures \cite{Ruili_Zhang_2016,Ruili_Zhang_2020}. The results, however, have not yet been related to the mechanism of vorticity wave interaction. 

In this work, we establish a link between the model of vorticity wave interaction and the analysis of $\mathcal{PT}$-symmetry breaking in shear flow instabilities. We show that the minimal dynamical system that describes coupling between two counter-propagating vorticity waves is a non-Hermitian, two-level system that exhibits a saddle-node exceptional point. The phase-locking of vorticity waves are then directly related to the Krein collision and breaking of $\mathcal{PT}$-symmetry through the exceptional point. The key parameter that leads the system to Krein collision and $\mathcal{PT}$-symmetry breaking is shown to be the ratio between frequency detuning and coupling strength of the vorticity waves. The analysis of $\mathcal{PT}$-symmetry breaking in shear flow instabilities is thus closely related to other two-level systems, such as coupled cavities or waveguides, that are extensively studied in non-Hermitian optics or photonics \cite{Bender2019,Ashida_2020}.

The most striking feature of non-Hermitian systems is the existence of exceptional points (EPs) \cite{Bender2019,Ashida_2020,Ding_2022,ARFM_2023}. EPs are spectral singularities on the complex eigenvalue plane where both the real and imaginary parts of eigenvalues are identical, and where the eigenvectors also coalesce and become identical. The eigenvectors are extremely nonorthogonal near an EP, and when a system operates around the EP, it becomes highly sensitive to perturbations of the system \cite{Bender2019, Ashida_2020, Ding_2022}. The transient dynamics such as power oscillations and amplifications near the EP are actively investigated in the context of non-Hermitian optics and photonics \cite{Power_oscillation_2008PRL,PRA2018,Makris_PRE2021}.  In this work, we apply the vorticity wave interaction approach to describe the critical behavior and transient dynamics near the EPs in shear flow instabilities. In particular, we demonstrate that the transition of phase dynamics from a phase-slip state to a phase-locking state \cite{pikovsky_2001,Strogatz_2015} corresponds to the spontaneous breaking of $\mathcal{PT}$-symmetry across the EP. The phase-slip dynamics near the EP are highly nonuniform in time and lead to non-modal, transient growth of perturbations \cite{Farrell_96,Nonmodal_2007,Nonmodal_2018,smyth_carpenter_2019} in the regime of unbroken $\mathcal{PT}$-symmetry. The phase-slip frequency $\Omega \propto |k-k_c|^{1/2}$ shares the same critical exponent with the phase rigidity of eigenvectors, which measures the nonorthogonality of system eigenvectors near the EP. We also extend the results to the dynamical system of multiple coupled vorticity waves, supported by a saw-tooth jet profile with multiple interfaces. It is shown that multiple EPs provide boundaries for $\mathcal{PT}$-symmetry breaking and onset of shear instabilities, characterized by the transition between phase-slip and phase-locking dynamics around the EPs.


\section{\label{sec: the model of vorticity wave interaction}The model of vorticity wave interaction}

\subsection{\label{sec: the general model}The general model}

Consider a two-dimentional, incompressible and inviscid shear flow without density stratification, where the background state is a mean flow in $y$ direction with profile $U(x)$. The velocity field is $\bm{v}=(v_x,v_y)=(-\partial \phi/\partial y, \partial \phi/\partial x)$ where $\phi$ is the stream function, and the vorticity $\bm{q}=\nabla \times \bm{v}$ of a 2D flow reduces to a scalar $ q=\nabla^2 \phi$. The vorticity equation reads $dq/dt=0$, where $d/dt=\partial/\partial t+\bm{v}\cdot\nabla$. To analyze  stability of the shear flow, denote perturbations of the background state as $v_x=\tilde{v}_x, v_y=U(x)+\tilde{v}_y, q=q_0+\tilde{q}$ and $\phi=\phi_0+\tilde{\phi}$, where the mean vorticity $q_0=\nabla^2 \phi_0=U'$, and mean vorticity gradient $q_0'=\partial q_0 /\partial x=U''$. (We use primes to denote $\partial_x$.) Then the linearized equation of perturbed vorticity is
\begin{equation}
    \frac{\partial \tilde{q}}{\partial t}+U\frac{\partial \tilde{q}}{\partial y}=U'' \frac{\partial \tilde{\phi}}{\partial y}.
     \label{linearized_vorticity}
\end{equation}

Eq.~(\ref{linearized_vorticity}) expresses the generation of vorticity perturbations by the background mean vorticity gradient, in terms of the source term $U'' \partial \tilde{\phi}/\partial y$. If all the perturbations are assumed to have normal mode form as $\tilde{\phi}(x,y,t)=\hat{\phi}(x)e^{ik( y- c t)}$ and $\tilde{q}(x,y,t)=\hat{q}(x)e^{ik(y-c t)}$, then the Rayleigh stability equation \cite{Drazin2002,Carpenter_2011}
\begin{equation}
    (U-c)(\hat{\phi}'' -k^2\hat{\phi})=U''\hat{\phi}
    \label{Rayleigh equation}
\end{equation}
is obtained for normal mode analysis of the problem. 

In our work, we take the non-modal approach \cite{Nonmodal_2007,Nonmodal_2018} and keep the time-dependent form of perturbations $\tilde{\phi}(x,y,t)=\hat{\phi}(x,t)e^{ik y}$ and $\tilde{q}(x,y,t)=\hat{q}(x,t)e^{ik y}$, with Fourier transform in $y$ direction, assuming that all the perturbations are monochromatic with certain wave number $k$. Then the linearized initial value problem is 
\begin{subequations}
\begin{gather}
    \frac{\partial \hat{q}}{\partial t}+i k U \hat{q}=ik U''\hat{\phi}, 
    \label{nonmodal_1}
    \\
    \hat{q}=\hat{\phi}''-k^2\hat{\phi}.
    \label{nonmodal_2}
    \end{gather}
\end{subequations}

We note that the assumption of monochromatic perturbations is adequate for the analysis of shear flow instabilities, since any perturbations in general can be expressed as a linear superposition of perturbations with different wave numbers.

Eq.~(\ref{nonmodal_2}) is the Poisson equation and can be solved as 
\begin{equation}
    \hat{\phi}(x,t)=\int G(x,x')\hat{q}(x',t)dx',
    \label{q_phi_Green}
\end{equation}
where the Green function kernel $G(x,x')$ describes a non-local $\hat{\phi}-\hat{q}$ coupling \cite{Mao_2022}. For boundary conditions $\hat{\phi}(\pm \infty)=0$, the kernel $G(x,x')=-e^{-k|x-x'|}/(2k)$. 

For a discretized, piecewise profile of background shear layer, neutrally stable vorticity waves are induced on each interface of vorticity jump, and the direction of intrinsic propagation is determined by the sign of the vorticity gradient \cite{Carpenter_2011}. In general, consider a piecewise shear layer with $N$ interfaces, 
\begin{equation}
    q_0'(x)  = U''(x)= \sum_{j=1}^{N}\Delta \bar{q}_j\delta(x-x_j),
\end{equation}
where $\Delta \bar{q}_j=q_0(x_j^+)-q_0(x_j^-)$ is the vorticity jump across the interface $x=x_j$.
The vorticity perturbations thus vanish at all locations except at the interfaces, \textit{i.e.}, 
\begin{equation}
    \hat{q}(x,t)=\sum_{j=1}^N q_j(t)\delta(x-x_j).
\end{equation}
Note that we have ignored the continuous spectrum part of the solution where initial vorticity perturbations are passively advected by the background flow and remain stable \cite{Heifetz_2005}.
Eq.~(\ref{q_phi_Green}) is then translated into 
\begin{equation}
\hat{\phi}(x,t)=-\frac{1}{2k} \sum_{j=1}^N q_j(t) e^{-k|x-x_j|}.
\label{discretized_poisson}
\end{equation}

Let $q_j(t)=Q_j(t)e^{i\theta_j(t)}$, 
Eq.~(\ref{nonmodal_1}) can be written in amplitude-phase form
\begin{subequations}
\begin{gather}
    \frac{dQ_i}{dt}= k \Delta \bar{q}_i \sum_{j \neq i} G_{ij} Q_j \sin\theta_{ij} , 
    \label{vortex_coupling_amplitude}
\\
    \frac{d\theta_i}{dt}=-\omega_i + k  \Delta \bar{q}_i \sum_{j \neq i} G_{ij} \frac{Q_j}{Q_i} \cos\theta_{ij},
    \label{vortex_coupling_phase}
    \end{gather}
\end{subequations}
where $\theta_{ij}=\theta_i-\theta_j$ and $G_{ij}=-e^{-k|x_i-x_j|}/(2k)$. 
The frequency of each individual mode
\begin{equation}
    \omega_i=k U_i + \frac{\Delta \bar{q}_i}{2}
    \label{vorticity wave frequency}
\end{equation}
includes both the intrinsic frequency $ \Delta \bar{q}_i/2$ and the Doppler shift by the \textit{in situ} shear flow $U_i$. Eq.~(\ref{vorticity wave frequency}) tells that each single interface of background vorticity jump $\Delta \bar{q}_i$ induces a neutrally stable vorticity wave, and the direction of phase speed $v_{ph}=- \Delta \bar{q}_i/(2k)$, in the reference frame moving with $U_i$, is determined by the sign of vorticity jump. 

Eq.~(\ref{vortex_coupling_amplitude}) and (\ref{vortex_coupling_phase}) describes a dynamical system that models vorticity wave interaction in a general shear layer with multiple interfaces. The system has a canonical Hamiltonian representation in action-angle form \cite{Heifetz_2009,heifetz_guha_2018}, with the Hamiltonian 
\begin{equation}
    H= \sum_i \omega_i A_i-\frac{1}{2} \sum_{j\neq i} G_{ij}Q_iQ_j\cos\theta_{ij} = -\sum_i A_i \dot{\theta}_i,
\end{equation}
where $A_i=Q_i^2/(2k\Delta \bar{q}_i)$ is the action of each vorticity wave. The canonical Hamiltonian equation follows as 
\begin{equation}
    \frac{\partial H}{\partial A_i}=-\dot{\theta}_i, \hspace{6mm} \frac{\partial H}{\partial \theta_i} =\dot{A}_i. 
\end{equation}

\begin{figure}[ht!]
\includegraphics[scale=0.6]{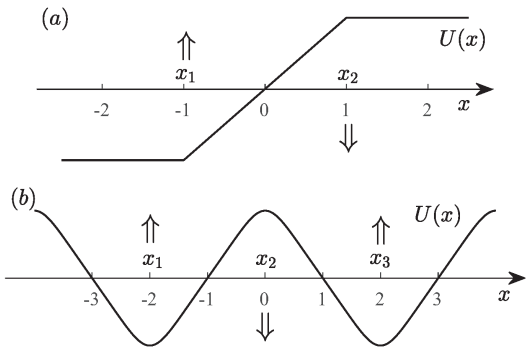}
\caption{\label{fig:profile} Profiles of the background shear layer and corresponding vorticity waves. (a) A piecewise shear layer profile that supports two counter-propagating vorticity waves, rescaled as $U(x_1)=-1$ and $U(x_2)=1$. (b) A saw-tooth jet profile with three interfaces, rescaled as $U(x_1)=U(x_3)=-1$ and $U(x_2)=1$. The arrows represent the direction of intrinsic propagation of vorticity waves.}
\end{figure}

To provide concrete examples of our main results, we consider minimal dynamical systems with two or three interfaces, as shown in Fig.~\ref{fig:profile}, in the remainder of the work. The background flow profile is nondimensionalized with the characteristic scale length $L_\ast =\Delta x/2$ and characteristic velocity $V_\ast=\Delta U/2$. Here, we denote $\Delta x=x_2-x_1$, $\Delta U=U(x_2)-U(x_1)$ for the two-interface shear layer shown in Fig.~\ref{fig:profile}(a), and $\Delta x=x_2-x_1=x_3-x_2$, $\Delta U=U(x_2)-U(x_{1,3})$ for the saw-tooth jet profile in Fig.~\ref{fig:profile}(b). 

\subsection{\label{sec: two waves}Interaction of two vorticity waves}

For a piecewise shear layer shown in Fig.~1(a), the interaction of two vorticity waves can be described by a minimal nonlinear dynamical system \cite{Heifetz_2019_PRE,guha_2014}, 
\begin{subequations}
\begin{gather}
    \dot{Q}_1=-\sigma Q_2 \sin \theta_{12}, \hspace{4mm} \dot{Q}_2=-\sigma Q_1 \sin \theta_{12}, 
    \label{two_vorticity_wave_system_a} 
    \\
    \dot{\theta}_1= -\omega_1-\sigma \frac{Q_2}{Q_1} \cos \theta_{12}, \hspace{4mm} \dot{\theta}_2= -\omega_2 + \sigma \frac{Q_1}{Q_2} \cos \theta_{12}
    \label{two_vorticity_wave_system_b}
\end{gather}
\label{two_vorticity_wave_system}
\end{subequations}
where $\omega_1=-k+1/2$, $\omega_2=k-1/2$,  and the coupling coefficient $\sigma=e^{-2k}/2$. The complex form of the dynamical system Eqs.~(\ref{two_vorticity_wave_system_a}) and (\ref{two_vorticity_wave_system_b}) is 
\begin{equation}
    \begin{split}
        \dot{q}_1=-i\omega_1 q_1-i\sigma q_2,
        \\
        \dot{q}_2=-i \omega_2 q_2+i \sigma q_1,
        \label{complex_two_vorticity_wave_system}
    \end{split}  
\end{equation}
or, in matrix form,
\begin{equation}
    \dot{\bm{q}}=\mathbf{A}\bm{q}, \hspace{8mm} \bm{q}=\begin{pmatrix} q_1\\q_2\end{pmatrix},
\end{equation}
where 
\begin{equation}
    \mathbf{A}=-i\mathbf{H}=-i\begin{pmatrix}
    \omega_1 & \sigma \\ -\sigma &  \omega_2
    \end{pmatrix}
\end{equation}
is a G-Hamiltonian matrix \cite{Yakubovich1975,Ruili_Zhang_2016,Ruili_Zhang_2020}. That is, $\mathbf{A}$ has the form of
$\mathbf{A}=-i \mathbf{G}^{-1}\mathbf{S}$, where $\mathbf{G}$ is a non-singular Hermitian matrix and $\mathbf{S}$ is a Hermitian matrix. Denote the frequency mismatch $\Delta \omega = \omega_1-\omega_2 = 1-2k$, then
\begin{equation}
    \mathbf{G}= \frac{1}{2k} 
    \begin{pmatrix}  1 & 0 \\ 0 & -1 \end{pmatrix},  \hspace{4mm}
    \mathbf{S}=\frac{1}{2k}  \begin{pmatrix}
         \frac{\Delta \omega}{2} & \sigma \\ \sigma & \frac{\Delta \omega}{2}
    \end{pmatrix}.
\end{equation}

The corresponding Hamiltonian is 
\begin{equation}
\begin{gathered}
    H(\bm{q})=\bm{q}^{\dag}\mathbf{S}\bm{q}\\
    =\omega_1 \frac{Q_1^2}{2k}-\omega_2 \frac{Q_2^2}{2k}+\frac{\sigma}{k}Q_1Q_2\cos \theta_{12},
\end{gathered}
\end{equation}
where $\bm{q}^{\dag}$ is the conjugate transpose of $\bm{q}$, and the complex canonical Hamiltonian equation has the form
\begin{equation}
    \dot{\bm{q}}=-i\mathbf{G}^{-1}\frac{\partial H}{\partial \bm{q}^*}.
    \label{complex_canonical_equation}
\end{equation}

The system conserves two constants of motion, one is the Hamiltonian $H(\bm{q})$ and the other is the total wave action $A(\bm{q})$. Define an indefinite inner product \cite{Kirillov2013} as 
\begin{equation}
    [\bm{x},\bm{y}]=(\mathbf{G}\bm{x},\bm{y})=\bm{y}^{\dag}\mathbf{G}\bm{x},
\end{equation}
then the total wave action can be written as
\begin{equation}
    A(\bm{q})=[\bm{q},\bm{q}]=\frac{Q_1^2}{2k}-\frac{Q_2^2}{2k}=A_1+A_2.
\end{equation}

The onset of Kelvin-Helmholtz instability can then be described via the dynamical system approach. Let $R_{12}=Q_1/Q_2$, and rewrite the dynamical system as
\begin{subequations}
\begin{gather}
    \dot{R}_{12}=\sigma (R_{12}^2-1)\sin \theta_{12},
    \label{dynamical_R_12}
    \\
    \dot{\theta}_{12}=-\Delta \omega -\sigma \big( R_{12}+\frac{1}{R_{12}}\big) \cos \theta_{12}.
    \label{dynamical_theta_12}
\end{gather}
\end{subequations}

\begin{figure}[ht!]
\includegraphics[scale=0.6]{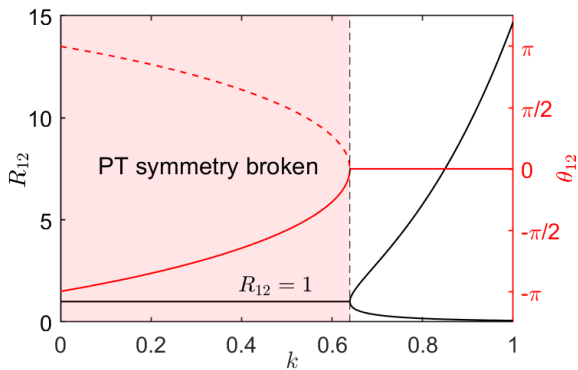}
\caption{\label{fig: fixed_point_two_wave} Parameter space of the two vorticity wave system and bifurcation of fixed points with the control parameter $k$. The black vertical dashed line is the critical $k=k_c\simeq0.64$. When $k \geq k_c$, the system has two neutrally stable fixed points, located at $\theta_{12}=0$, with unbroken $\mathcal{PT}$-symmetry. When $0<k<k_c$, the system has a pair of stable (solid red line, $-\pi<\theta_{12}<0$) and unstable (dashed red line, $0<\theta_{12}<\pi$) fixed points, and $\mathcal{PT}$-symmetry is spontaneously broken.}
\end{figure}

The fixed points of the system require either 
\begin{equation}
    \sin \theta_{12}=0, \hspace{4mm} R_{12}+\frac{1}{R_{12}} =\frac{\Delta \omega}{\sigma \cos \theta_{12}},
    \label{fixed_point_unbroken}
\end{equation}
where the two vorticity waves remain neutrally stable, or
\begin{equation}
    R_{12}=1, \hspace{4mm} \cos \theta_{12}=-\frac{\Delta \omega}{2\sigma},
    \label{fixed_point_broken}
\end{equation}
where the two vorticity waves are phase-locked. 
The control parameter of the bifurcation of fixed points is the ratio between frequency detuning and coupling strength of the vorticity waves,
\begin{equation}
    \mu=-\frac{\Delta \omega}{2\sigma}=(2k-1)e^{2k}.
\end{equation}
Note that $\mu$ is a monotonously increasing function of $k$ when $k\geq0$, so that the parameter space for instability onset $-1<\mu<1$ is equivalent to $0<k<k_c \simeq 0.64$, as shown in Fig.~\ref{fig: fixed_point_two_wave}. 

Therefore, the onset of shear instability corresponds to a bifurcation of fixed points from two neutral centers to a pair of stable and unstable nodes \cite{Heifetz_2019_PRE}. When $k > k_c$, the phase detuning $\Delta \omega$ dominates over coupling strength $\sigma$ between the two vorticity waves, and the system has two neutrally stable fixed points. When $0<k< k_c$, the phase-locking of two vorticity waves corresponds to a pair of stable and unstable fixed points, depending on the sign of $\sin\theta_{12}$. The $-\pi<\theta_{12}<0$ fixed point is stable as a dynamical sink of phase plane trajectories, which triggers mutual growth and instability onset, as shown in the solid red line of Fig.~\ref{fig: fixed_point_two_wave}; the $0<\theta_{12}<\pi$ fixed point is unstable as a source of phase plane trajectories, corresponding to unphysical, mutual damping of initial perturbations, as shown in the dashed red line of Fig.~\ref{fig: fixed_point_two_wave} (Also see Fig.~4 and 6 in Ref.~\cite{Heifetz_2019_PRE}).  

The nondimensionalized wave number of perturbations $k$ is the key parameter throughout our work. Here, we note that for the dimensional system, the parameter space for instability onset is $0<k\Delta x <1.28$. The physical meaning of the key parameter is then explained as the wavelength of vorticity waves compared to the distance between the vorticity jump interfaces \cite{Carpenter_2011}. When $k\Delta x \gg 1.28$, the distance between adjacent counter-propagating vorticity waves are too large compared to their wavelength, so that the waves do not feel each other's presence and remains neutrally stable, as if they were isolated. When $k\Delta x$ is varied to go below the critical value, $k\Delta x<1.28$, interaction of the vorticity waves becomes stronger and the waves become phase-locked.

\section{Krein collision and $\mathcal{PT}$-symmetry breaking\label{subsec: Krein collision}}
The model of vorticity wave interaction provides a clear physical description and a non-modal approach to understand the mechanism of Kelvin-Helmholtz instabilities. Here, we show that it also provides a natural and simple testbed for the rich non-Hermitian physics in shear flow instabilities. In particular, it has been shown in Ref.~\cite{Hong_Qin_2019_PoP,Fu_2020_NJP,Tomos_David_2023} that Kelvin-Helmholtz instability is the result of spontaneous $\mathcal{PT}$-symmetry breaking, and we demonstrate in this section that the result can be well explained and appreciated through the mechanism of vorticity wave interaction. 

To proceed, we first list the following theorems \cite{Ruili_Zhang_2020} that relate the concepts of G-Hamiltonian matrix, pseudo-Hermiticity, and $\mathcal{PT}$-symmetry, without giving proofs. The proofs can be found in Ref. \cite{Ruili_Zhang_2020}.

\textbf{Theorem 1}: \textit{For a finite-dimensional system $\dot{\bm{q}}=\mathbf{A}\bm{q}=-i\mathbf{H}\bm{q}$, $\mathbf{H}$ is pseudo-Hermitian if and only if $\mathbf{A}$ is a G-Hamiltonian matrix. }

\textbf{Theorem 2}: \textit{For a finite-dimensional system, a $\mathcal{PT}$-symmetric Hamiltonian $\mathbf{H}$ is necessarily pseudo-Hermitian.}

Theorem 1 and Theorem 2 guarantee that for finite-dimensional systems, when a Hamiltonian operator $\mathbf{H}$ is $\mathcal{PT}$-symmetric, it is necessarily pseudo-Hermitian and the corresponding matrix $\mathbf{A}=-i\mathbf{H}$ is G-Hamiltonian. It follows that $\mathcal{PT}$-symmetric operators constitute a subclass of pseudo-Hermitian operators \cite{Ashida_2020}.

For the two-vorticity wave dynamical system, the parity operator $\mathcal{P}$ and the time reversal operator $\mathcal{T}$ acting on vorticity fluctuation variables $\{q_1(t),q_2(t)\}$ are
\begin{equation}
    \mathcal{P}=\begin{pmatrix} -1 & 0 \\ 0 & -1 \end{pmatrix} \hspace{4mm} \text{and} \hspace{4mm} \mathcal{T}=\mathcal{K},
\end{equation}
satisfying $\mathcal{P}^2=\mathcal{T}^2=\mathcal{I}_{2\times 2}$, where $\mathcal{K}$ is the complex-conjugate operator and $\mathcal{I}$ is the identity operator. One can readily see that the Hamiltonian operator $\mathbf{H}$ is $\mathcal{PT}$-symmetric, \textit{i.e.}, $\mathcal{PT}\mathbf{H}-\mathbf{H}\mathcal{PT}=0$. It is also straightforward to verify that $\mathbf{H}$ is pseudo-Hermitian with $\mathbf{G}\mathbf{H}=\mathbf{H}^{\dag}\mathbf{G}$.

Now, it is equivalent to consider the eigensystem of pseudo-Hermitian operator $\mathbf{H}$, $\mathbf{H}\bm{u}=E\bm{u}$, or the eigensystem of G-Hamiltonian operator $\mathbf{A}$, $\mathbf{A}\bm{u}=\lambda\bm{u}$, satisfying $\lambda=-iE$. Therefore, the systematically developed theory of G-Hamiltonian systems \cite{Yakubovich1975} can be readily applied. The eigenvalues of a G-Hamiltonian matrix are categorized as follows: a purely imaginary eigenvalue $\lambda$ of a G-Hamiltonian matrix $\mathbf{A}$ is called definite with positive Krein signature if $[\bm{u},\bm{u}]>0$ for any eigenvector $\bm{u}$ in its eigensubspace, and with negative Krein signature if $[\bm{u},\bm{u}]<0$. We recall the following properties of the eigenvalues of a G-hamiltonian matrix:

\textbf{Theorem 3}: \textit{The eigenvalues of a G-Hamiltonian matrix are symmetric with respect to the imaginary axis.}

\textbf{Theorem 4}: \textit{For a G-Hamiltonian matrix $\mathbf{A}$, if the matrix $\mathbf{G}$ has $p$ positive eigenvalues and $q$ negative eigenvalues, then $\mathbf{A}$ has $p$ eigenvalues with positive Krein signature and $q$ eigenvalues with negative Krein signature.}

\textbf{Theorem 5}: (\textit{Krein-Gel'fand-Lidskii theorem}) \textit{A G-Hamiltonian system $\dot{\bm{q}}=\mathbf{A}\bm{q}=-i\mathbf{H}\bm{q}$ is strongly stable (\textit{i.e.}, the stability of the system is preserved by any infinitesimal deformation of the Hamiltonian $\mathbf{H}$) if and only if all eigenvalues of $\mathbf{A}$ lie on the imaginary axis and are definite.}

The proofs can be found in Ref.~\cite{Yakubovich1975}. An immediate corollary is that, for the G-Hamiltonian matrix $\mathbf{A}$ in our system, the only route to become unstable is through the overlap between two eigenvalues with opposite Krein signatures on the imaginary axis, known as the Krein collision \cite{Yakubovich1975,Kirillov2013,Kirillov2014,Ruili_Zhang_2016,Ruili_Zhang_2020}. Note that Theorem 3, 4 and 5 have been applied in Ref. \cite{Ruili_Zhang_2016} to study the two-stream instability in plasmas. It is shown there that the physical interpretation of the Krein signature is the sign of the action of the corresponding eigenmode, and the onset of two-stream instabilities is illustrated as the Krein collision between a positive-action mode and a negative-action mode \cite{Ruili_Zhang_2016}. We demonstrate that it is also the case in our system of shear instabilities. 

To see this, for eigenvector $\bm{u}$, from $\mathbf{H}=\mathbf{G}^{-1}\mathbf{S}$ and the corresponding Hamiltonian $H(\bm{u})=\bm{u}^{\dag}\mathbf{S}\bm{u}$, we obtain
\begin{equation}
    [\bm{u},\bm{u}]=\frac{H(\bm{u})}{E},
    \label{action_u}
\end{equation}
which is the ratio between energy and eigenfrequency of the eigenmode.
Recall that the Krein signature $\kappa (\lambda)$ is defined as
\begin{equation}
    \kappa (\lambda) =\text{sign}[\bm{u},\bm{u}].
\end{equation}
Therefore, Eq.~(\ref{action_u}) shows that the physical interpretation of the Krein signature is the sign of the action of the corresponding eigenmode. In our system, the eigenvalues of $\mathbf{H}$ are 
\begin{equation}
    E_{1,2}=\pm \frac{\sqrt{\Delta \omega^2 -4 \sigma^2}}{2} 
\end{equation}
and the eigenvectors
\begin{equation}
    \bm{u}_{1,2}=\begin{pmatrix}
        \displaystyle -\frac{\Delta \omega}{2\sigma} \mp \frac{\sqrt{\Delta \omega^2 -4 \sigma^2}}{2\sigma} \vspace{2mm} \\  1
    \end{pmatrix}.
\end{equation}
Denote the discriminant of the characteristic polynomial of $\mathbf{H}$ as $D=\Delta \omega^2 -4 \sigma^2=(2k-1)^2-e^{-4k}$. For $k>k_c$, $D>0$ and both eigenvalues $E_{1,2}$ are real, suggesting that the system has unbroken $\mathcal{PT}$-symmetry. The eigenvectors $\bm{u}_i=(u_{i1} \hspace{2mm} u_{i2})^T$ ($i=1,2$) of $\mathbf{H}$ are also eigenvectors of $\mathcal{PT}$, with
\begin{equation}
    \mathcal{PT} \begin{pmatrix}
        u_{i1}\\u_{i2}
    \end{pmatrix}= \begin{pmatrix}
        -u_{i1}^*\\-u_{i2}^*
    \end{pmatrix}= - \begin{pmatrix}
        u_{i1}\\u_{i2}
    \end{pmatrix}.
    \label{PT symmetry analysis}
\end{equation}

For $0<k<k_c$, $D<0$ and $\mathbf{H}$ has a pair of complex conjugate eigenvalues $E_{1,2}=\pm i\sigma\sqrt{1-(\Delta\omega/2\sigma)^2}=-i \sigma \sin \theta_{12}$, suggesting the spontaneously breaking of $\mathcal{PT}$-symmetry of the system, as shown in Fig.~\ref{fig: fixed_point_two_wave}. Note that the eigenvectors $\bm{u}_{1,2}=(e^{i\theta_{12}} \hspace{2mm} 1)^T$ of $\mathbf{H}$ are not eigenvectors of $\mathcal{PT}$ anymore since $\mathcal{PT}$-symmetry is broken.

\begin{figure}[ht!]
\includegraphics[scale=0.6]{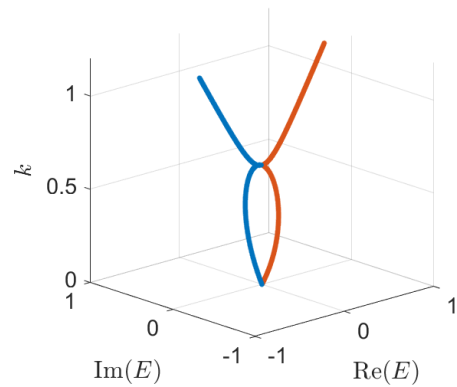}
\caption{\label{fig: EP_two_wave} The eigenvalues $E_{1}(k)$ (red) and $E_{2}(k)$ (blue) in the complex (Re($E$),Im($E$),$k$) space. The saddle-node exceptional point locates at $k=k_c$.}
\end{figure}

The boundary of this change is marked by the spectral singularity at $k=k_c$, known as the exceptional point (EP). At the EP, $D=0$ and both real and imaginary parts of $E_{1,2}$ are identical. Therefore, the EP is a saddle point in the complex (Re($E$),Im($E$),$k$) space as shown in Fig.~\ref{fig: EP_two_wave}, and the onset of instability is via bifurcation through the EP. Note that the eigenvectors $\bm{u}_{1,2}$ are also identical at the EP, which distinguishes the EP from diabolic points in Hermitian systems, where the eigenvalues are degenerate but the eigenvectors remain complete and orthogonal. The unique features of the exceptional point is a focus of our work and will be addressed in detail in section \ref{sec: exceptional point}.

\begin{figure}[ht!]
\includegraphics[scale=0.56]{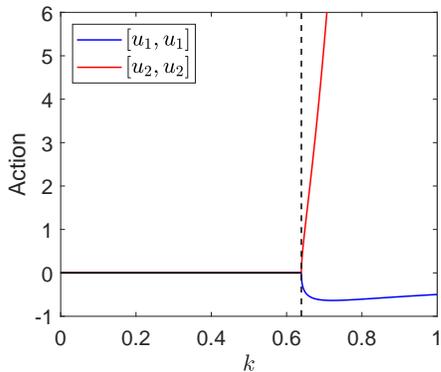}
\caption{\label{fig: wave_action} The indefinite inner products (actions) of eigenmodes, $[\bm{u}_1,\bm{u}_1]$ and $[\bm{u}_2,\bm{u}_2]$. The Krein collision requires a negative-action eigenmode and a positive-action eigenmode to collide at $k=k_c$.}
\end{figure}

Now, the indefinite inner products of the eigenmodes are
\begin{equation*}
    [\bm{u}_{1,2}, \bm{u}_{1,2}]= \begin{cases}
     \displaystyle \frac{-1+e^{4k}(1-2k\pm \sqrt{D})^2}{2k}, & k>k_c
    \\ 0, & 0<k<k_c
    \end{cases}
\end{equation*} 
 as shown in Fig.~\ref{fig: wave_action}. When $k>k_c$, the eigenvalues $\lambda_{1,2}=-iE_{1,2}$ lie on the imaginary axis and the system is spectrally stable. The Krein signatures are $\kappa(\lambda_1)=\text{sign}[\bm{u}_1,\bm{u}_1]=-1$ and $\kappa(\lambda_2)=\text{sign}[\bm{u}_2,\bm{u}_2]=1$, corresponding to a negative-action mode and a positive-action mode. When $k\rightarrow k_c$, the two eigenvalues $\lambda_{1,2}$ collide and the two eigenmodes also coalesce, with $\bm{u}_1=\bm{u}_2$. That is the Krein collision between a negative-action mode and a positive-action mode. When $0<k<k_c$, the two eigenvalues move off the imaginary axis, and the actions of both eigenmodes vanish. The Krein collision of eigenvalues (and corresponding eigenmodes) thus illustrates the breaking of $\mathcal{PT}$-symmetry and the onset of Kelvin-Helmholtz instabilities. 
\begin{figure*}[ht!]
    \centering
    \includegraphics[scale=1]{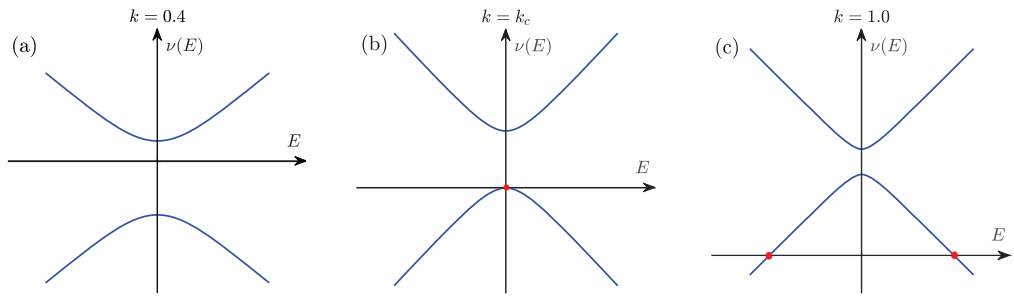}
    \caption{Graphical interpretation of the Krein signature, with two eigenvalue branches of the pencil $\mathcal{L}$ at (a) $k=0.4$, (b) $k=k_c\simeq 0.6392$ and (c) $k=1$. The zeros of the eigenvalue branches correspond to purely imaginary eigenvalues of $\mathbf{A}$.}
    \label{fig: graphic}
\end{figure*}

The Krein signatures can also be obtained by a graphical method \cite{SIAM_2014}. Consider a Hermitian linear pencil $\mathcal{L}=\mathbf{S}-E\mathbf{G}$, and solve the eigenvalue problem
\begin{equation}
    \mathcal{L}(E)\bm{u}(E)=(\mathbf{S}-E\mathbf{G})\bm{u}(E)=\nu(E)\bm{u}(E)
    \label{linear_pencil}
\end{equation}
parameterized by $E \in \mathbb{R}$, where $\nu=\nu(E)$ is called an eigenvalue branch of $\mathcal{L}$. The intersection points of $\nu(E)$ with $\nu=0$ axis then correspond to real eigenvalues of $\mathbf{H}$. 
Differentiate Eq. (\ref{linear_pencil}) at $E=E_0$ and $\nu(E_0)=0$, we obtain
\begin{equation}
    (\mathbf{S}-E_0\mathbf{G})\bm{u}'(E_0)=\nu'(E_0)\bm{u}(E_0)+\mathbf{G} \bm{u}(E_0).
\end{equation}
Taking the inner product with $\bm{u}(E_0)$ then gives
\begin{equation}
    \nu'(E_0)(\bm{u},\bm{u})=-(\mathbf{G}\bm{u},\bm{u}),
\end{equation}
so that
\begin{equation}
    \kappa(\lambda_0)=-\text{sign}\big[ \nu'(E_0)\big].
\end{equation}
Therefore, we can determine the Krein signatures of the imaginary spectrum of $\mathbf{A}$ simply by looking at the sign of the slope of the eigenvalue branch $\nu=\nu(E)$ at the intersection points. In our system, we plot
\begin{equation}
    \nu_{1,2}(E)=\frac{\Delta \omega}{4k}\pm \frac{\sqrt{E^2+\sigma^2}}{2k}
\end{equation}
for specific choices of the control parameter $k$ in Fig.~\ref{fig: graphic}. When $k>k_c$, there are two intersection points with opposite signs of the slope, corresponding to two imaginary eigenvalues of $\mathbf{A}$ with opposite Krein signatures, and the system is spectrally stable. When $k\rightarrow k_c$, the pair of adjacent intersection points collide at $E=0$, leading to the Krein collision. When $k<k_c$, the eigenvalue branches $\nu=\nu_{1,2}(E)$ have no zeros, implying spectral instability of the system.
\begin{figure*}[ht!]
    \centering
    \includegraphics[scale=0.52]{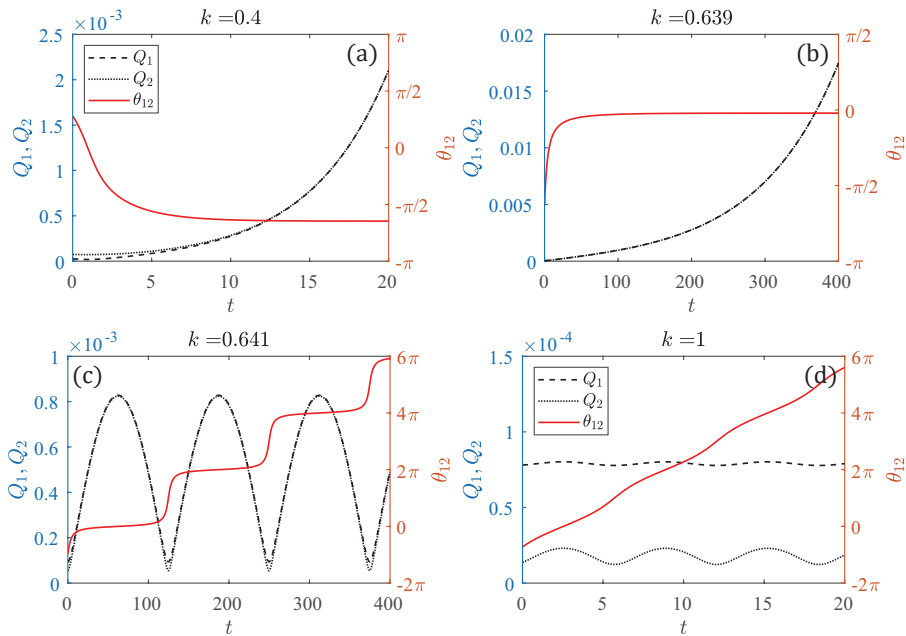}
    \caption{The amplitude and phase dynamics of two coupled vorticity waves, calculated from Eq.~(\ref{two_vorticity_wave_system_a}) and (\ref{two_vorticity_wave_system_b}). The initial perturbations are random with amplitude $\sim 10^{-5}$. The control parameter of the system is (a) $k=0.4$, (b) $k=0.639\sim k_c^-$, (c) $k=0.641\sim k_c^+$, and (d) $k=1$. }
    \label{fig: dynamics}
\end{figure*}

In general, we have applied the model of vorticity wave interaction to demonstrate the Krein collision and $\mathcal{PT}$-symmetry breaking in shear flow instabilities. The key parameter that leads the system to spontaneously $\mathcal{PT}$-symmetry breaking is the ratio $\mu=-\frac{\Delta \omega}{2\sigma}$, which measures the competition between vorticity wave frequency detuning and coupling strength. We further note that the fixed points of the vorticity wave dynamical system in Eq.~(\ref{fixed_point_unbroken}) and (\ref{fixed_point_broken}) are equivalent to the normal mode solutions  $\bm{u}_1 e^{-iE_1 t}$ and $\bm{u}_2 e^{-iE_2 t}$ of the problem \cite{Heifetz_2019_PRE}, since the minimal nonlinear dynamical system Eq.~(\ref{two_vorticity_wave_system}) is equivalent to its linear, complex form Eq.~(\ref{complex_two_vorticity_wave_system}). The amplitude and phase dynamics of the two vorticity waves for arbitrary initial condition can be obtained from superposition of normal modes,
\begin{equation}
    \bm{q}(t)=c_1\bm{u}_1 e^{-iE_1 t}+c_2\bm{u}_2 e^{-iE_2 t},
\end{equation}
as shown in Fig.~\ref{fig: dynamics}. The parameter space $0<k<k_c$ corresponds to phase-locking dynamics and exponential growth of initial perturbations, as shown in Fig.~\ref{fig: dynamics}(a) and \ref{fig: dynamics}(b), when the coupling strength dominates over phase detuning. The parameter space $k>k_c$ corresponds to phase-slip dynamics and transient growth of initial perturbations, as shown in Fig.~\ref{fig: dynamics}(c), when the coupling strength is weak compared with the phase detuning. Note that when $k\gg k_c$, the vorticity waves decouple and remain neutrally stable as shown in \ref{fig: dynamics}(d), as the coupling strength decays exponentially with decreasing wavelength or increasing distance between the waves.

We can thus compare the results above to the $\mathcal{PT}$-symmetry analysis in Ref.~\cite{Fu_2020_NJP}. It is shown in \cite{Fu_2020_NJP} that the phase difference between $\hat{v}_x(x,t)$ and $\hat{v}_y(x,t)$ are locked to $\pi/2$ when the system is stable with unbroken $\mathcal{PT}$-symmetry, and become arbitrary and spatially dependent when the system is unstable with broken $\mathcal{PT}$-symmetry. In our work, the system converges to the exponentially growing fixed point for the unstable regime and the normal mode solution
\begin{equation}
    \hat{\phi}(x,t)=-\left( \frac {e^{i\theta_{12}}e^{-k|x+1|}+e^{-k|x-1|}}{2k} \right) e^{-\sigma \sin\theta_{12}t}
\end{equation}
with $-\pi<\theta_{12}<0$ has a rather complicated mode structure. Therefore, $\hat{v}_x=-ik\hat{\phi}$ and $\hat{v}_y=\hat{\phi}'$ are phase-locked in time, but with arbitrary and spatially dependent phase differences. For the stable regime with unbroken $\mathcal{PT}$-symmetry, we have $\theta_{12}=0$ and $\theta_{1}=\theta_{2} \equiv \theta(t)$ whenever $\bm{q}(t)$ converges to one of the normal modes (fixed points). We then obtain $\hat{\phi}(x,t)=|\hat{\phi}(x)|e^{i[\theta(t)+\pi]}$, and therefore, $\hat{v}_x=-ik\hat{\phi}$ and $\hat{v}_y=\hat{\phi}'$ are phase-locked to $\pi/2$, which is consistent with Ref.~\cite{Fu_2020_NJP}. 

However, we emphasize that when the $\mathcal{PT}$-symmetry is unbroken, the fixed points are neutrally stable and the physical solution for arbitrary initial condition, being the superposition of normal modes, does not need to converge to any of them. The nonorthogonality of eigenvectors near the EP then leads to transient phase-slip dynamics, which is an intriguing feature of the EP and beyond the scope of \cite{Fu_2020_NJP}. We shall concentrate on the critical behavior around the EP in the following section.

\section{\label{sec: exceptional point}Critical behavior around the exceptional point}

One of the most striking features of non-Hermitian systems is the existence of exceptional points (EPs). EPs are branch singularities on the complex eigenvalue plane where both the real and imaginary parts of eigenvalues are identical \cite{Ding_2022}. In Hermitian systems, the critical point where the eigenvalues degenerate are called diabolic points. At a diabolic point, the eigenvalues are identical but the eigenvectors remain complete and orthogonal. In non-Hermitian systems, the eigenvectors also coalesce and become identical at an EP. This nonorthogonality of eigenvectors near the EP is a unique feature of non-Hermitian systems. The transient dynamics around the EP is actively investigated in the context of non-Hermitian optics and photonics, where power oscillations or amplifications have been observed \cite{Power_oscillation_2008PRL}. These observations are closely related to the transient growth in fluid dynamics \cite{smyth_carpenter_2019}, and reveal a generic property of the EP that when a system operates around an EP, it becomes highly sensitive to perturbations of the system \cite{Bender2019,Ashida_2020,Ding_2022}. In this section, we demonstrate this characteristic of EP from the perspective of vorticity wave resonance in shear flow instabilities, and describe this transient behavior as a transition between phase-locking and phase-slip dynamics. 

Denote the left and right eigenvectors of $\mathbf{H}$ as $\langle u^L|$ and $|u^R\rangle$, and the right eigenvectors of $\mathbf{H}^\dag$ as $\bm{v}$, \textit{i.e.} $\mathbf{H}^\dag \bm{v} =E^*\bm{v}$, then we obtain $|u^R\rangle =\bm{u}$ and $\langle u^L|=\bm{v}^\dag$. For the two-vorticity wave system,
\begin{equation*}
    \langle u_{1,2}^L|=
     \Big(
        \displaystyle \frac{\Delta \omega}{2\sigma} \pm \frac{\sqrt{\Delta \omega^2 -4 \sigma^2}}{2\sigma} \hspace{2mm} 1
    \Big),
\end{equation*}
and the eigenvectors of the non-Hermitian system are biorthonormal, with $\langle u_{1}^L|u_{2}^R\rangle=\langle u_{2}^L|u_{1}^R\rangle=0$. A quantitative measure for the nonorthogonality of the eigenvectors is the phase rigidity \cite{Rotter_2009,Rotter_2017_PRA,Ding_2022}, defined as
\begin{equation}
    r=\frac{\langle u^L|u^R\rangle}{\langle u^R|u^R\rangle}.
\end{equation}

\begin{figure}[ht!]
\includegraphics[scale=0.56]{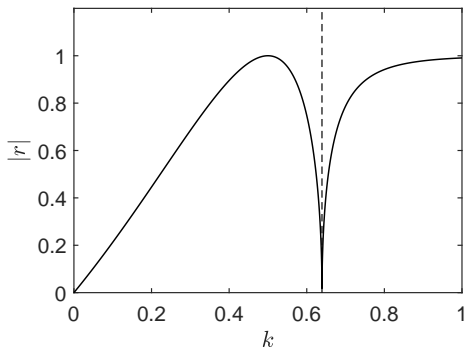}
\caption{\label{fig: phase_rigidity} Plot of the phase rigidity $|r|$ as function of $k$. At the EP, $|r|=0$ and the eigenvectors are identical. In the vicinity of the EP, $|r|\propto |k-k_c|^{1/2}$ and the eigenvectors are extremely nonorthogonal. When $k\gg k_c$, the coupling is weak and $|r|\rightarrow 1$. Note that at $k=0.5$, the frequency of two vorticity waves are naturally equal regardless of coupling and $|r|\simeq 1$.}
\end{figure}

In Hermitian systems, the right eigenvectors are orthogonal and the phase rigidity $r=1$. In non-Hermitian systems, the right eigenvectors of different states are skewed instead of orthogonal, and the phase rigidity is parameter dependent, with $|r|<1$. Approaching the EP, the two eigenvectors coalesce and $r\rightarrow 0$. For the two-vorticity wave system, 
\begin{equation}
    |r_{1,2}|=\begin{cases}
    \displaystyle \sqrt{1-\frac{4\sigma^2}{\Delta \omega^2}}, & k>k_c \vspace{4mm} \\
       \displaystyle \sqrt{1-\frac{\Delta \omega^2}{4\sigma^2}}, &0<k<k_c
    \end{cases}
\end{equation}
as plotted in Fig.~\ref{fig: phase_rigidity}. Expand the phase rigidity near the EP, and we obtain $|r_{1,2}|\propto |k-k_c|^{1/2}$. In the vicinity of the EP, the phase rigidity quantifies the splitting of eigenvectors \cite{Ding_2022}, which follows a square-root dependence similar to the splitting of eigenvalues, $\delta E=|E_1-E_2|\propto |k-k_c|^{1/2}$.  

The critical exponent $s=1/2$ is associated with critical behavior of the vorticity wave dynamical system near the EP. Note that when the system crosses through the EP from $k=k_c^+$ to $k=k_c^-$, a transition of phase dynamics occurs from a phase-slip state (Fig.~\ref{fig: dynamics}(c)) to a phase-locking state (Fig.~\ref{fig: dynamics}(b)). The transition is captured by
\begin{equation}
    \dot{\theta}_{12}=-\Delta \omega -2\sigma \cos \theta_{12},
    \label{Adler}
\end{equation}
which is obtained from Eq. (\ref{dynamical_theta_12}) in the vicinity of the EP, where $R_{12}\rightarrow 1$. Eq.~(\ref{Adler}) is the Adler equation \cite{pikovsky_2001,Strogatz_2015}, which contains both phase-locking and phase-slip solutions.

The phase-slip dynamics of $\theta_{12}(t)$ are highly nonuniform in time in the vicinity of the EP, and correspond to transient growth of initial perturbations \cite{Heifetz_2005,guha_2014,guha_2017} in the regime of unbroken $\mathcal{PT}$-symmetry. Denote the phase-slip period $T$ as the time required for $\theta_{12}$ to jump by $2\pi$. For most of the time in the period, the dynamical phase $\theta_{12}$ is getting through the bottleneck around multiples of $2\pi$, predicted by the neutrally stable fixed point when $k>k_c$. More precisely, $\theta_{12}$ is around $2n\pi^-$ in the first half of the period, which corresponds to transient growth; and around $2n\pi^+$ in the second half of the period, which corresponds to transient damping. A phase-slip period then ends with a short interval of $2\pi$ phase jump, so that the initial perturbations endure periodic oscillations of non-modal growth and damping. When $k\gg k_c$ (Fig.~\ref{fig: dynamics}(d)), the phase-slip period $T\rightarrow 2\pi/\Delta \omega$ and $\theta_{1,2}(t)\propto t$, so that normal modes are recovered and transient growth vanishes.

The phase-slip period $T$ near the EP is estimated as
\begin{equation}
    T=\int_0^{2\pi}\frac{d\theta_{12}}{-\Delta \omega -2 \sigma \cos \theta_{12}}=\frac{2\pi}{\sqrt{\Delta \omega^2-4\sigma^2}}, 
    \label{phase-slip period}
\end{equation}
and the phase slip frequency $\Omega=\sqrt{\Delta \omega^2-4\sigma^2}$.
\begin{figure}[ht!]
\includegraphics[scale=0.56]{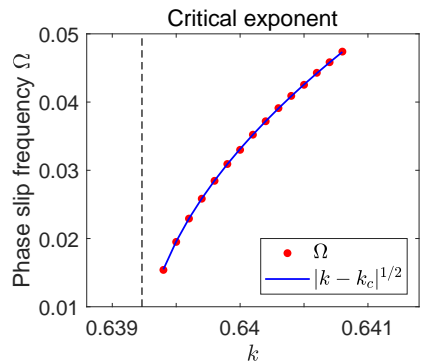}
\caption{\label{fig: phase_slip_frequency} The critical exponent of phase slip frequency $\Omega$ near the EP is $s=1/2$. The red dots are numerically calculated from Eq. (\ref{two_vorticity_wave_system_a}) and (\ref{two_vorticity_wave_system_b}), and the solid blue line is $\propto|k-k_c|^{1/2}$.}
\end{figure}
Therefore, the phase-slip frequency satisfies $\Omega \propto |k-k_c|^{1/2}$ near the EP, following a square-root scaling law, as shown in Fig.~\ref{fig: phase_slip_frequency}. The results are consistent with the square-root dependence of the phase rigidity, showing that when the eigenvectors are extremely nonorthogonal, the phase dynamics become extremely non-modal and highly nonuniform in time accordingly. Therefore, applying the vorticity wave interaction approach, the transition of phase dynamics provides a useful description of the critical behavior around the EP.

The transient growth of perturbations near the EP
\begin{equation}
    \frac{Q(t)}{Q(0)}=\exp \left[ -\sigma \int_0^t \sin \theta_{12}(t) dt\right]
\end{equation}
can also be estimated for the first half-period as
\begin{equation}
    \frac{Q(\frac{T}{2})}{Q(0)}=\exp \left[ -\sigma \int_{\cos \theta_0}^1 \frac{d \cos \theta_{12}}{\Delta \omega+2\sigma \cos \theta_{12}} \right] = \sqrt{\frac{\mu -\cos \theta_0}{\mu-1 }},
\end{equation}
where $\theta_0=\theta_{12}(0)$. The optimal amplification factor $G$ for transient growth is then obtained when $\cos \theta_0=-1$, as
\begin{equation}
    G=\sqrt{\frac{\mu+1}{\mu-1}}.
    \label{optimal}
\end{equation}
We have thus recovered the optimal transient growth rate Eq.~(\ref{optimal}) that is consistent with previous results \cite{Heifetz_2005,guha_2014}. See the Appendix for details. We conclude that the transient growth of perturbations, along with the transition from phase-locking to phase-slip dynamics, describe the essential characteristics of an EP in shear flow instabilities.

\section{Interaction of three vorticity waves}
In this section, we extend the results above to the interaction of multiple vorticity waves, supported by a saw-tooth jet profile with multiple interfaces \cite{guha_2017}. The multiple EPs provide boundaries for $\mathcal{PT}$-symmetry breaking and onset of shear instabilities, characterized by the transition between phase-slip and phase-locking dynamics around the EPs. We thus show that the $\mathcal{PT}$-symmetry analysis using the vorticity wave interaction approach is not limited to a two-level system, but applicable to more complicated flow profiles relevant in various settings, such as zonal jets in the atmosphere and oceans \cite{McIntyre2008}. 

\subsection{Dynamical system analysis}

Consider a saw-tooth jet shear layer shown in Fig.~1(b). The interaction of three vorticity waves can be described by a complex nonlinear dynamical system \cite{guha_2017}, 
\begin{equation}
    \begin{split}
        \dot{q}_1&=-i\omega_1 q_1-i\sigma_1 q_2-i\sigma_2 q_3,
        \\
        \dot{q}_2&=-i \omega_2 q_2+i \sigma_1 q_1 +i \sigma_1 q_3,
        \\
        \dot{q}_3&=-i \omega_3 q_3-i \sigma_1 q_2-i \sigma_2 q_1,
        \label{complex_three_vorticity_wave_system}
    \end{split}  
\end{equation}
where $\omega_1=-k+1$, $\omega_2=k-1$, $\omega_3=-k+1$, and the coupling coefficients $\sigma_1=e^{-2k}$, $\sigma_2=e^{-4k}$. 
In matrix form, we have
\begin{equation}
    \dot{\bm{q}}=\mathbf{A}\bm{q}, \hspace{8mm} \bm{q}=\begin{pmatrix} q_1&q_2&q_3\end{pmatrix}^T,
\end{equation}
where 
\begin{equation}
    \mathbf{A}=-i\mathbf{H}=-i\begin{pmatrix}
    \omega_1 & \sigma_1 & \sigma_2 \\ -\sigma_1 &  \omega_2 &-\sigma_1 \\ \sigma_2 & \sigma_1 & \omega_3
    \end{pmatrix}
\end{equation}
is a G-Hamiltonian matrix that has the form of
$\mathbf{A}=-i \mathbf{G}^{-1}\mathbf{S}$. Denote the frequency mismatch $\Delta \omega = \omega_1-\omega_2 =\omega_3-\omega_2= 2-2k$, then the non-singular Hermitian matrix $\mathbf{G}$ and the Hermitian matrix $\mathbf{S}$ reads
\begin{equation}
    \mathbf{G}= \frac{1}{4k} 
    \begin{pmatrix}  1 & 0 & 0\\ 0 & -1 & 0 \\ 0 & 0 & 1 \end{pmatrix},  \hspace{4mm}
    \mathbf{S}=\frac{1}{4k} \begin{pmatrix}
        \frac{\Delta \omega}{2} & \sigma_1 & \sigma_2 \\ \sigma_1 & \frac{\Delta \omega}{2} & \sigma_1 \\ \sigma_2 & \sigma_1 & \frac{\Delta \omega}{2}
    \end{pmatrix}.
\end{equation}
\begin{figure}[ht!]
    \centering
    \includegraphics[scale=0.58]{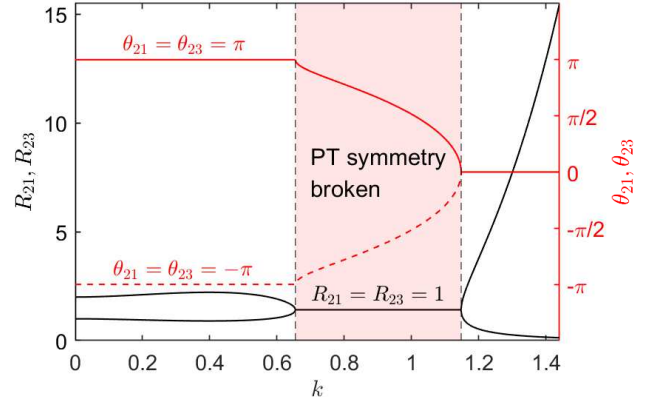}
    \caption{ Parameter space of the three vorticity wave system and bifurcation of fixed points with the control parameter $k$. The black vertical dashed lines are the critical $k=k_{c1}\simeq 0.6545$ and $k=k_{c2}\simeq 1.1475$. When $k \leq k_{c1}$ or $k \geq k_{c2}$, the system has two neutrally stable fixed points, located at $\theta_{21}=\theta_{23}=\pm \pi$ or $\theta_{21}=\theta_{23}=0$, with unbroken $\mathcal{PT}$-symmetry. When $k_{c1}<k<k_{c2}$, the system has a pair of stable (solid red line, $0<\theta_{21}=\theta_{23}<\pi$) and unstable (dashed red line, $-\pi<\theta_{21}=\theta_{23}<0$) fixed points, and $\mathcal{PT}$-symmetry is spontaneously broken. The three vorticity waves become phase-locked and lead to shear instabilities.}
    \label{fig: fixed_point_three_wave}
\end{figure}

The corresponding Hamiltonian is 
\begin{equation}
\begin{gathered}
    H(\bm{q})=\bm{q}^{\dag}\mathbf{S}\bm{q}
    =\omega_1 \frac{Q_1^2}{4k}-\omega_2 \frac{Q_2^2}{4k}+\omega_3 \frac{Q_3^2}{4k}\\
    +\frac{\sigma_1}{2k}Q_1Q_2\cos \theta_{12}+\frac{\sigma_1}{2k}Q_2Q_3\cos \theta_{23}+\frac{\sigma_2}{2k}Q_3Q_1\cos \theta_{31},
\end{gathered}
\end{equation}
where $\bm{q}^{\dag}$ is the conjugate transpose of $\bm{q}$. The canonical equation still has the form of Eq.~(\ref{complex_canonical_equation}), and the total wave action 
\begin{equation}
    A(\bm{q})=[\bm{q},\bm{q}]=\frac{Q_1^2}{4k}-\frac{Q_2^2}{4k}+\frac{Q_3^2}{4k}=A_1+A_2+A_3.
\end{equation}

Let amplitude ratios $R_{ij}=Q_i/Q_j$, and rewrite the dynamical system (see Eqs.~(3.7)--(3.10) in Ref.~\cite{guha_2017}) as 
\begin{widetext}
\begin{subequations}
    \begin{gather}
        \dot{R}_{21}=(1-R_{21}^2)\sin \theta_{21}e^{-2k}+\frac{R_{21}}{R_{23}}\sin \theta_{23} e^{-2k} -\frac{R_{21}^2}{R_{23}}\sin \theta_{31} e^{-4k}, \\
        \dot{R}_{23}=(1-R_{23}^2)\sin \theta_{23}e^{-2k}+\frac{R_{23}}{R_{21}}\sin \theta_{21} e^{-2k} +\frac{R_{23}^2}{R_{21}}\sin \theta_{31} e^{-4k}, \\
        \dot{\theta}_{21}=-2k+2+\big( R_{21}+\frac{1}{R_{21}}\big)\cos \theta_{21} e^{-2k} 
        +\frac{1}{R_{23}}\cos \theta_{23} e^{-2k} + \frac{R_{21}}{R_{23}}\cos \theta_{31} e^{-4k}, \\
        \dot{\theta}_{23}=-2k+2+\big( R_{23}+\frac{1}{R_{23}}\big)\cos \theta_{23} e^{-2k} 
        +\frac{1}{R_{21}}\cos \theta_{21} e^{-2k} + \frac{R_{23}}{R_{21}}\cos \theta_{31} e^{-4k}. 
    \end{gather}
\end{subequations}
\end{widetext}

Solving $\dot{R}_{21}=0$ and $\dot{R}_{23}=0$ then gives either 
\begin{equation}
    \text{Case 1:} \hspace{4mm} \frac{1}{R_{21}^2}+\frac{1}{R_{23}^2}=1
\end{equation}
or
\begin{equation}
    \text{Case 2:} \hspace{4mm}  \frac{\sin \theta_{21}}{R_{21}}+\frac{\sin \theta_{23}}{R_{23}}=0.
\end{equation}

For Case 1, $\dot{\theta}_{21}=0$ and $\dot{\theta}_{23}=0$ give the relation
\begin{equation}
    \cos \theta_{31}=\frac{R_{23}R_{21}}{2}\geq 1,
\end{equation}
so that $\theta_{31}=0$ and $R_{21}=R_{23}=\sqrt{2}$,
\begin{equation}
\cos \theta_{21}= \cos \theta_{23} =-\frac{\Delta \omega+\sigma_2}{2\sqrt{2} \sigma_1}=\frac{(2k-2)e^{2k}-e^{-2k}}{2\sqrt{2}},
\end{equation}
which requires $k_{c1}<k<k_{c2}$, where $k_{c1}\simeq 0.6545$ and $k_{c2} \simeq 1.1475$. When $0<\theta_{21}=\theta_{23}<\pi$, the fixed point is stable and the phase-locking of three vorticity waves trigger shear instabilities. When $-\pi<\theta_{21}=\theta_{23}<0$, the fixed point is unstable, corresponding to decay of initial perturbations. 

For Case 2, we obtain $\sin \theta_{21}=\sin \theta_{23}=0$, so that the fixed points require either $\theta_{21}=\theta_{23}=\pm \pi$ and
\begin{equation}
    R_{21}=R_{23}=\frac{\Delta \omega+\sigma_2}{2\sigma_1} \pm \frac{\sqrt{(\Delta \omega+\sigma_2)^2-8\sigma_1^2}}{2\sigma_1},
\end{equation}
when $k<k_{c1}$; or $\theta_{21}=\theta_{23}=0$ and
\begin{equation}
    R_{21}=R_{23}=-\frac{\Delta \omega+\sigma_2}{2\sigma_1} \pm \frac{\sqrt{(\Delta \omega+\sigma_2)^2-8\sigma_1^2}}{2\sigma_1},
\end{equation}
when $k>k_{c2}$. The fixed points are neutrally stable when $k<k_{c1}$ or $k>k_{c2}$, as shown in the bifurcation diagram of fixed points in Fig.~\ref{fig: fixed_point_three_wave}. (See also Fig.~2 in Ref. \cite{guha_2017}.) Note that a singular case is ignored \cite{guha_2017} when $(2k-1)+e^{-4k}=0$ and $k\simeq 0.3984$, where $\theta_{31}=\pm \pi$ and the eigenvalue branches $E_1(k)$ and $E_3(k)$ intersect.

We also note that for the dimensional system, the parameter space for instability onset is $2k_{c1} <k\Delta x<2k_{c2} $, \textit{i.e.}, the wavelength of vorticity waves needs to be comparable to the distance between adjacent vorticity interfaces, as a compromise between coupling strength and frequency detuning.

\subsection{Exceptional points and $\mathcal{PT}$-symmetry breaking}
The Hamiltonian $\mathbf{H}$ of the three vorticity wave system is also $\mathcal{PT}$-symmetric, with parity operator $\mathcal{P}=-\mathcal{I}_{3\times3}$ and time reversal operator $\mathcal{T}=\mathcal{K}$. The eigenvalues of $\mathbf{H}$ are
\begin{equation}
    E_{1,2}=\frac{\sigma_2}{2} \pm \frac{\sqrt{(\Delta \omega+\sigma_2)^2-8\sigma_1^2}}{2} 
\end{equation}
and $E_3=\frac{\Delta \omega}{2}-\sigma_2$. The eigenvectors are
\begin{equation}
    \bm{u}_{1,2}=\begin{pmatrix}
        1 & \displaystyle -\frac{\Delta \omega+\sigma_2}{2\sigma_1} \pm \frac{\sqrt{(\Delta \omega+\sigma_2)^2-8\sigma_1^2}}{2\sigma_1} &1
    \end{pmatrix}^T
\end{equation}
and $\bm{u}_3=\begin{pmatrix}
        -1&0&1
    \end{pmatrix}^T.$
The discriminant of the characteristic polynomial of $\mathbf{H}$ is
\begin{equation}
    D=4(\sigma_1^2+\sigma_2^2-\Delta \omega \sigma_2)^2\big[(\Delta \omega+\sigma_2)^2-8\sigma_1^2\big].
\end{equation}
\begin{figure}[h]
\includegraphics[scale=0.6]{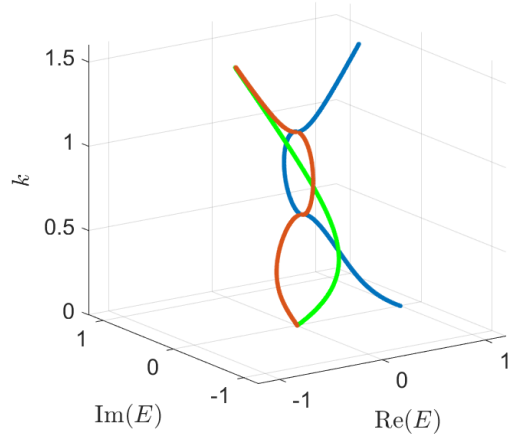}
\caption{\label{fig: EP_three_wave} The eigenvalues $E_{1}(k)$ (blue), $E_{2}(k)$ (red) and $E_{3}(k)$ (green) in (Re($E$),Im($E$),$k$) space. The two saddle-node EPs locate at $k=k_{c1}$ and $k=k_{c2}$.}  
\end{figure}
\begin{figure*}[ht!]
    \centering
    \includegraphics[scale=0.5]{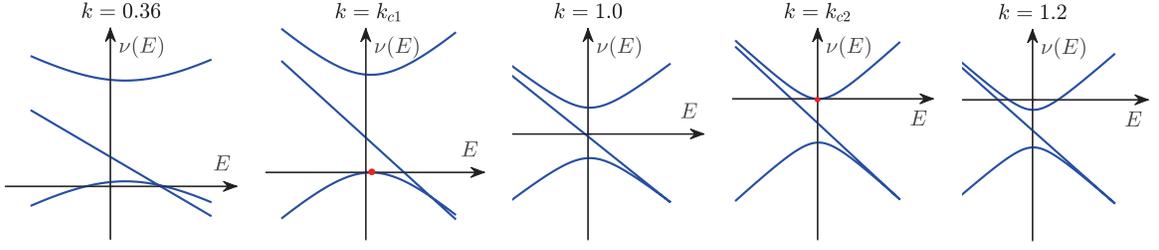}
    \caption{Graphical interpretation of the Krein signature, with three eigenvalue branches of the pencil $\mathcal{L}$ at (a) $k=0.36$, (b) $k=k_{c1}\simeq 0.6545$, (c) $k=1$, (d) $k=k_{c2}\simeq 1.1475$ and (e) $k=1.2$. The zeros of the eigenvalue branches correspond to purely imaginary eigenvalues of $\mathbf{A}$.}
    \label{fig: graphic_three}
\end{figure*}
\begin{figure*}[ht!]
    \centering
    \includegraphics[scale=0.36]{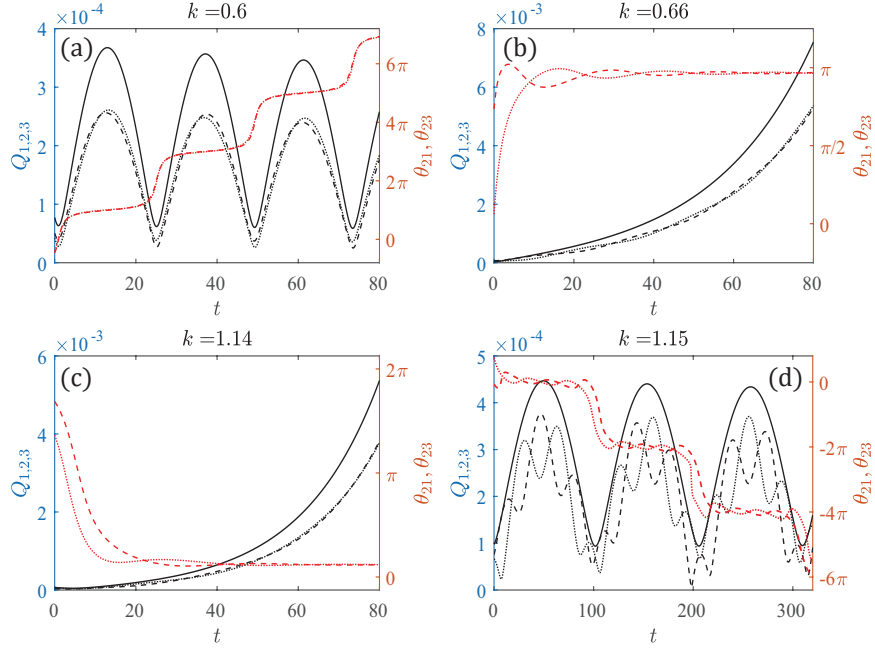}
    \caption{The amplitude and phase dynamics of three coupled vorticity waves, with random initial perturbations $\sim 10^{-5}$. The control parameter of the system is (a) $k=0.6$, (b) $k=0.66$, (c) $k=1.14$, and (d) $k=1.15$. The transition of phase dynamics between a phase-slip state and a phase-locking state exhibits around $k=k_{c1}$ and $k=k_{c2}$.}
    \label{fig: dynamics_three}
\end{figure*}

Therefore, when $0<k<k_{c1}$ or $k>k_{c2}$, $D \geq 0$ and the eigenvalues $E_{1,2,3}$ are real, the system has unbroken $\mathcal{PT}$-symmetry; when $k_{c1}<k<k_{c2}$, $D<0$ and $\mathbf{H}$ has a pair of pure imaginary eigenvalues $E_{1,2}$, implying the spontaneously breaking of $\mathcal{PT}$-symmetry. The boundaries of this change is marked by two saddle-node EPs at $k=k_{c1}$ and $k=k_{c2}$, as shown in Fig.~\ref{fig: EP_three_wave}.

The eigenvalue branches $\nu(E)$ of the Hermitian pencil $\mathcal{L}=\mathbf{S}-E\mathbf{G}$ are then
plotted in Fig.~\ref{fig: graphic_three}. The pair of imaginary eigenvalues of $\mathbf{A}$ with opposite Krein signatures $\lambda_{1,2}=-iE_{1,2}$ collide at $k=k_{c1}^-$ and $k=k_{c2}^+$, and the G-Hamiltonian system becomes spectrally unstable when $k_{c1}<k<k_{c2}$. The transition from phase-locking to phase-slip dynamics is shown in Fig.~\ref{fig: dynamics_three}, with the phase-slip frequency
\begin{equation}
    \Omega=\sqrt{(\Delta \omega +\sigma_2)^2-8\sigma_1^2}.
\end{equation}
The critical exponent is then $s=1/2$ as the phase-slip frequency $\Omega \propto |k-k_{c1,2}|^{1/2}$ near the EPs.
\begin{figure}[ht!]
\includegraphics[scale=0.6]{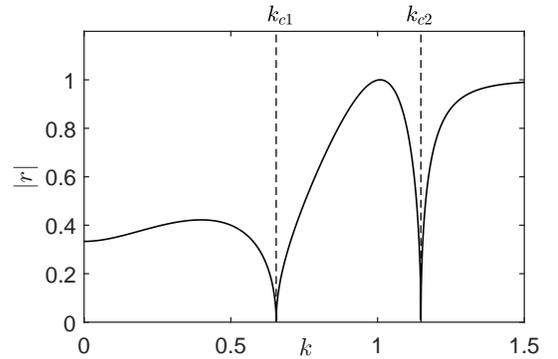}
\caption{\label{fig: phase_rigidity_three} Plot of the phase rigidity $|r|$ as function of $k$ in three vorticity wave system. Around the EPs $k=k_{c1}$ and $k=k_{c2}$, $|r|\propto|k-k_{c}|^{1/2}$ with the critical exponent $1/2$.}
\end{figure}

The phase rigidity of system eigenvectors is 
\begin{equation}
    |r_{1,2}|=\begin{cases}
    \displaystyle \sqrt{1-\frac{8\sigma_1^2}{(\Delta \omega +\sigma_2)^2}}, & k<k_{c1} \text{ or } k>k_{c2} \vspace{1mm}\\
       \displaystyle \sqrt{1-\frac{(\Delta \omega +\sigma_2)^2}{8\sigma_1^2}}, & k_{c1}<k<k_{c2} 
    \end{cases}
\end{equation}
as plotted in Fig.~\ref{fig: phase_rigidity_three}, with $|r|\propto |k-k_{c1,2}|^{1/2}$ near the EPs. 
Therefore, we conclude that the square-root dependence is a universal characteristic of the EPs in vorticity wave systems.

\section{Summary and discussion}
In this work, we relate the model of vorticity wave interaction to the analysis of $\mathcal{PT}$-symmetry breaking in shear flow instabilities. We show that the minimal dynamical system that describes coupling of vorticity waves is a non-Hermitian system with saddle-node EPs. The mechanism of spatial coupling and phase-locking of vorticity waves provides a clear physical interpretation to explain the spontaneous breaking of $\mathcal{PT}$-symmetry in shear flow instabilities. The key parameter that leads to Krein collision and $\mathcal{PT}$-symmetry breaking is shown to be the ratio between frequency detuning and coupling strength of the vorticity waves. 

A highlight of this work is to describe the critical behavior near the EPs as a transition between phase-slip and phase-locking dynamics of the vorticity waves. We show that this transition of phase dynamics corresponds to the spontaneous $\mathcal{PT}$-symmetry breaking and onset of shear instabilities. In particular, the phase-slip dynamics are highly nonuniform in time, corresponding to the extreme nonorthogonality of eigenvectors near the EPs, measured by the phase rigidity of system eigenvectors. The phase-slip dynamics lead to non-modal, transient growth of perturbations near the EP in the regime of unbroken $\mathcal{PT}$-symmetry. The phase-slip frequency shares the same critical exponent $1/2$ with the phase rigidity of system eigenvectors, indicating the square-root dependence as a universal characteristic of EPs in vorticity wave systems. The results are extended to the dynamical system of multiple coupled vorticity waves, where multiple EPs provide boundaries for $\mathcal{PT}$-symmetry breaking and onset of shear instabilities.

In general, we have shown that the framework of vorticity wave interaction is naturally related to non-Hermitian physics and $\mathcal{PT}$-symmetry-breaking bifurcations. Therefore, we anticipate further applications of this framework not limited to instabilities in shear flows, but also in analysis of instabilities in other continuous media, such as drift wave instabilities \cite{Qin_PRE_2021, Mao_2022} and magnetohydrodynamic instabilities \cite{heifetz_2015,YZhang_2020} in plasmas. 
\begin{center}

\end{center}
\begin{acknowledgments}

We acknowledge useful discussions with J. Q. Li, J. Wang, Z. J. Mao and Y. Zhang. This work was supported by the National Natural Science Foundation of China with grant Nos. 12075013, 12275071 and U1967206.
\end{acknowledgments}

\appendix*

\section{\label{appendix}Transient Growth Analysis}

In this Appendix, we discuss alternative approaches to obtain the transient growth rate \cite{Farrell_96,Heifetz_2005,guha_2014,PRA2018,Makris_PRE2021} in the regime of unbroken $\mathcal{PT}$-symmetry, as comparison with the results estimated from phase-slip dynamics.

Given initial condition $\bm{q}(0)=\bm{q}_0$, the general solution of the initial value problem is $\bm{q}(t)=e^{\mathbf{A}t}\bm{q}_0$, and the propagator matrix of the dynamical system,
    $e^{\mathbf{A}t}=e^{-i\mathbf{H}t}$,
can be obtained from superposition of normal modes,
\begin{equation}
\begin{cases}
    \bm{q}(t)=c_1\bm{u}_1 e^{-iE_1 t}+c_2\bm{u}_2 e^{-iE_2t},\\
    \bm{q}(0)=c_1\bm{u}_1+c_2\bm{u}_2.
    \end{cases}
    \label{solve c}
\end{equation}
The coefficients $c_1, c_2$ are readily solved from Eq. (\ref{solve c}), and one gets
\begin{equation}
    e^{\mathbf{A}t}=
    \begin{pmatrix}
        \cos{E_1 t}-i\frac{\Delta\omega}{2E_1}\sin{E_1 t}&-i\frac{\sigma}{E_1} \sin E_1 t \\
      i\frac{\sigma}{E_1} \sin E_1 t&\cos{E_1 t}+i\frac{\Delta\omega}{2E_1}\sin{E_1 t} 
    \end{pmatrix},
\end{equation}
where $E_1=\sqrt{(\Delta \omega/2)^2-\sigma^2}$. The amplification factor of transient growth under certain initial condition is then 
\begin{equation}
\begin{split}
    G^2(t)&=\frac{(\bm{q}(t),\bm{q}(t))}{(\bm{q}_0,\bm{q}_0)}=\frac{(e^{\mathbf{A}^{\dag}t}e^{\mathbf{A}t} \bm{q}_0,\bm{q}_0)}{(\bm{q}_0,\bm{q}_0)}
    \\&=\frac{|q_1(t)|^2+|q_2(t)|^2}{|q_1(0)|^2+|q_2(0)|^2}.
\end{split}
\end{equation}

The optimal growth can be obtained either using the singular value decomposition (SVD) approach \cite{Farrell_96,Heifetz_2005} or directly calculating $G^2(t)$ \cite{PRA2018}. The SVD of $e^{\mathbf{A}t}$ has the form of $e^{\mathbf{A}t}=\mathbf{U}\Sigma \mathbf{V}^{\dag}$, where the columns of $\mathbf{U}$ and $\mathbf{V}$ are eigenvectors of $e^{\mathbf{A}t}e^{\mathbf{A}^{\dag}t}$ and $e^{\mathbf{A}^{\dag}t}e^{\mathbf{A}t}$, separately. In the regime of unbroken $\mathcal{PT}$-symmetry, the singular value matrix is
\begin{equation}
    \Sigma=\begin{pmatrix}
        g&0\\0&g^{-1}
    \end{pmatrix}, \hspace{4mm} g=\sqrt{\frac{\mu-\cos\theta_0}{\mu+\cos \theta_0}},
\end{equation}
so that the optimal growth factor is
\begin{equation}
    G=\sqrt{\frac{\mu+1}{\mu-1}}
\end{equation}
with the optimal initial condition $Q_1(0)=Q_2(0)$ and $\cos \theta_0 =-1$. 
Direct estimation of $G^2(t)$ also recovers the optimal initial condition, and the transient oscillation dynamics under such condition are given by
\begin{equation}
    G^2(t)= 1 + \frac{2}{\mu-1} \sin^2 E_1t.
\end{equation}
The results are consistent with Eq.~(\ref{optimal}) in the main text, where the oscillation period $T=\pi/E_1$ is just the phase-slip period estimated from Eq.~(\ref{phase-slip period}).


\bibliography{apssamp}

\end{document}